\documentclass[fleqn,usenatbib]{mnras}

\usepackage{newtxtext,newtxmath}

\usepackage[T1]{fontenc}

\DeclareRobustCommand{\VAN}[3]{#2}
\let\VANthebibliography\thebibliography
\def\thebibliography{\DeclareRobustCommand{\VAN}[3]{##3}\VANthebibliography}

\usepackage{graphicx}	
\usepackage{amsmath}	
\usepackage{hyperref}

\title[The Formation of NGC~1052-DF9]{Keck Spectroscopy of NGC~1052-DF9: Stellar Populations in the Context of the NGC~1052 Group}

\author[J. S. Gannon et al.]{
Jonah S. Gannon$^{1,2}$\thanks{E-mail: jgannon@swin.edu.au},
Maria Luisa Buzzo$^{1,2}$,
Anna Ferr\'e-Mateu$^{3,4}$, 
Duncan A. Forbes$^{1,2}$,
\newauthor{Jean P. Brodie$^{1,2,5}$, 
and Aaron J. Romanowsky$^{6,5}$}
\\
$^{1}$ Centre for Astrophysics and Supercomputing, Swinburne University, John Street, Hawthorn VIC 3122, Australia
\\
$^{2}$ ARC Centre of Excellence for All Sky Astrophysics in 3 Dimensions (ASTRO 3D)
\\
$^{3}$ Instituto de Astrof\'isica de Canarias, Calle V\'ia L\'actea S/N, E-38205, La Laguna, Tenerife, Spain
\\ 
$^{4}$ Departamento de Astrofísica, Universidad de La Laguna, 38206, La Laguna (S.C. Tenerife), Spain
\\
$^{5}$ Department of Astronomy \& Astrophysics, University of California Santa Cruz, 1156 High Street, Santa Cruz, CA 95064, USA
\\
$^{6}$ Department of Physics and Astronomy, San Jos\'e State University, One Washington Square, San Jose, CA 95192, USA
\\
}

\date{Accepted XXX. Received YYY; in original form ZZZ}

\pubyear{2023}

\begin{document}
\label{firstpage}
\pagerange{\pageref{firstpage}--\pageref{lastpage}}
\maketitle

\begin{abstract}
In this study, we use Keck/KCWI spectroscopy to measure the age, metallicity and recessional velocity of NGC~1052-DF9 (DF9), a dwarf galaxy in the NGC~1052 group. We compare these properties to those of two other galaxies in the group, NGC~1052-DF2 and NGC~1052-DF4, which have low dark matter content. The three galaxies are proposed constituents of a trail of galaxies recently hypothesised to have formed as part of a ``bullet dwarf'' collision. We show that the ages and total metallicities of the three galaxies are within uncertainties of one another which may be expected if they share a related formation pathway. However, the recessional velocity we recover for DF9 (1680 $\pm$ 10 km s$^{-1}$) is higher than predicted for a linearly projected interpretation of the ``bullet dwarf'' trail. DF9 is then either not part of the trail or the correlation of galaxy velocities along the trail is not linear in 2D projection due to their 3D geometry. After examining other proposed formation pathways for the galaxies, none provide a wholly satisfactory explanation for all of their known properties. We conclude further work is required to understand the formation of this interesting group of galaxies.
\end{abstract}




\section{Introduction}
The galaxies NGC~1052-DF2 (hereafter, DF2)\footnote{DF2 was first discovered by \citet{Fosbury1978} although it was not named until more recently \citep{Karachentsev2000}.} and NGC~1052-DF4 (hereafter, DF4)\footnote{For a full summary of additional designations for both DF2 and DF4, see Table 1 of \citet{Roman2021b}.} in the NGC~1052 group have been highly controversial due to claims of their dark matter deficient nature \citep{vanDokkum2018, vanDokkum2019, Danieli2019} and their over-luminous globular cluster (GC) populations \citep{vanDokkum2018, vanDokkum2019, Shen2020}. Many attempts have been made to describe the formation of each galaxy individually, not considering these galaxies may be related. These include tidal stripping from a massive object \citep{Ogiya2018, Ogiya2021, Nusser2020, Montes2020, Maccio2021, Jackson2021, Moreno2022}; their formation as tidal dwarf galaxies from stripped HI gas \citep{vanDokkum2018, Fensch2018}; or as the result of extreme interstellar medium (ISM) conditions from e.g., a high redshift merger \citep{TrujilloGomez2021}. If the formation pathway of these two galaxies is a rare occurrence, it would be an unlikely coincidence to find two in the same galaxy group (although see \citealp{Moreno2022}). Furthermore, some of these formation scenarios (e.g., tidal stripping) do not explain the population of over-luminous GCs around each galaxy \citep{Shen2020}.

In a separate scenario, the formation of DF2, DF4 and up to 9 other galaxies in the NGC~1052 group may be the result of a high-speed galaxy collision \citep{Silk2019, Shin2020, Lee2021, vanDokkum2022}. Briefly, \citet[hereafter vD+22]{vanDokkum2022} identified a near-linear ``trail'' of galaxies in projection and suggested these may be the result of a high-speed galaxy collision $\sim$~8 Gyr ago. This collision separates off the existing stars, GCs and dark matter from the gas, which later collapses to form new galaxies and GCs. Such a scenario then leaves the two progenitor galaxies dark matter dominated and stripped of gas at either end of a trail of newly formed dark matter-free galaxies (see further in extended figure 1, vD+22). They dubbed this proposal the ``bullet dwarf'' scenario due to similarities with a miniaturised version of the bullet cluster.

Key predictions from the ``bullet dwarf'' scenario are that the newly formed galaxies and their constituents (e.g., GCs) should have:
\begin{enumerate}
    \item Similar ages and metallicities due to their formation from the same stripped gas.
    \item Recessional velocities that are correlated with their on-sky position. However, we note that this correlation is not expected to be linear in 2D projection for most 3D geometries in the system (vD+22). 
    \item An extreme lack of dark matter, reflective of their formation from gas stripped from a dark matter halo.
\end{enumerate}

The trail galaxies DF2 and DF4 have already been tested for each of these three requirements. Both have been found to be dark matter deficient \citep{vanDokkum2018, vanDokkum2019, Danieli2019, Emsellem2019, Montes2020, Keim2022}. Additionally, DF2 and DF4 have had their stellar population parameters (i.e., age and metallicity) measured from either spectroscopy (DF2; \citealp{Fensch2018, RuizLara2019}) or spectral energy distribution (SED) fitting (DF2/DF4; \citealp{Buzzo2022}). The ages and metallicities for both galaxies are within uncertainties of one another \citep{Buzzo2022}. Finally, both DF2 and DF4 have measured recessional velocities \citep{Danieli2019, vanDokkum2019} that place initial constraints on the geometry of the collision (vD+22).

%

In this work, we study the next brightest galaxy located along this bullet trail, NGC~1052-DF9 (hereafter, DF9). After DF2 and DF4, DF9 is the only galaxy remaining on the trail bright enough to be observable with current spectrographs in reasonable integration times. We note that DF9 was excluded by vD+22 as a member of the objectively selected trail as it was not in the catalogue of \citet{Roman2021b} due to its relative brightness. However, vD+22 still noted it may be a trail member. It is also one of only two other galaxies on the trail that has a GC system and is nucleated (\citealp{Buzzo2023}). It is therefore an important test of the bullet trail hypothesis. Throughout this work, we assume a distance to DF9 of 20 Mpc ($m-M=31.5$) which is the same distance assumed in the related work of \citet{Buzzo2023} along with the published works of \citet{vanDokkum2018c} and \citet{Shen2020} which concern DF2/DF4. We note that there has been some controversy about this assumed distance to these dwarf galaxies (see e.g., \citealp{Trujillo2019}), but recent deep \textit{Hubble Space Telescope} imaging supports our assumption \citep{Shen2021}. Magnitudes are in the AB system.


\begin{figure}
    \centering
    \includegraphics[width = 0.48 \textwidth]{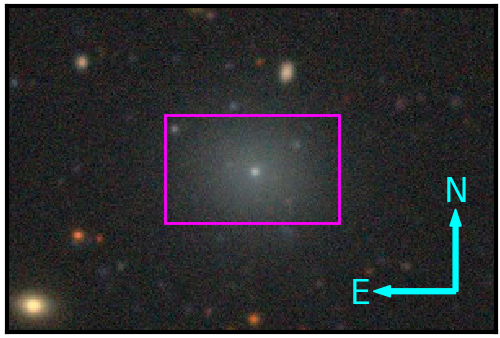}
    \caption{A 90 arcsec $\times$~60 arcsec (8.4 $\times$~5.6 kpc at NGC~1052 distance), DECaLS cutout centred on DF9. The magenta rectangle indicates the FOV of our KCWI observations. North and east are as indicated (cyan arrows). We identify the central compact source as the nucleus of DF9. }
    \label{fig:fig1}
\end{figure}

\section{Keck Cosmic Web Imager Data} \label{sec:data}


The Keck Cosmic Web Imager (KCWI) data used in this work was observed on the night 2022, January 29 as part of program N195 (PI: Romanowsky). Conditions were clear with 0.9'' seeing in which a single 1200s exposure was observed. The data were observed using the Large slicer with the BL grating and cover a wavelength range of 3554\AA~to 5574\AA. A spectral resolution of $R=785$~($\sigma_{\rm Inst} = 160$ km s$^{-1}$)~at 4500\AA~was measured on the calibration arc files. This resolution was degraded from expected instrument performance due to software errors during instrument configuration. 

These data were reduced using the standard KCWI data reduction pipeline (\citealp{Morrissey2018}) and were subsequently cropped per \citet{Gannon2020}. Two spectra were extracted from the resulting data cube: 1) a spectrum of the nucleus and 2) a galaxy spectrum excluding the central nucleus. The nucleus spectrum was extracted using a $4\times1$ spaxel ($1.17\times1.35$ arcsec) box with the remainder of the slicer, excluding a 1 spaxel buffer around the extraction box, as background (i.e., both sky and galaxy). It has a final signal-to-noise ratio of 13 \AA$^{-1}$. The galaxy spectrum was extracted using a $41 \times 13$ spaxel ($11.95\times17.65$ arcsec) box with the outskirts of the data cube as background. Two compact sources, including the nucleus, were removed from both the galaxy and background regions. It has a final signal-to-noise ratio of 14 \AA$^{-1}$. We display a DECaLS stamp centred on DF9 (\url{https://www.legacysurvey.org/viewer}), along with KCWI pointing, in Figure \ref{fig:fig1}.

\begin{table}
    \centering
    \begin{tabular}{llll}
    \hline
    RA (J2000) & 02:40:07.01 \\
    Dec (J2000) & $-$08:13:44.4 \\ 
    $m_g$ [mag] & 17.17 \\
    $M_{\star}$ [$\mathrm{M_{\odot}}$] & $1.31\times10^{8}$ \\
    $\mu_{g}(0)$ [mag arcsec$^{-2}$] & 23.81 \\
    $R_{\rm e}$ {[}arcsec{]} & 11.1  \\
    $R_{\rm e}$ [kpc] & 1.08  \\
     $V_{\rm r}$ {[}km s$^{-1}${]} & 1680 (10) \\
     D [Mpc] & 20 \\ 
     \hline
    \end{tabular}
    \caption{The basic properties of NCG~1052-DF9. From top to bottom, the entries are: 1) Right Ascension (J2000); 2) Declination (J2000); 3) $g-$band apparent magnitude ($m_g$); 4) total stellar mass ($M_{\star}$); 5) central $g-$band surface brightness ($\mu_{g}(0)$); 6) half-light radius in arcseconds; 7) half-light radius in kpc at the assumed 20 Mpc distance ($R_{\rm e}$); 8) recessional velocity ($V_{\rm r}$) with \texttt{pPXF} given uncertainty and 9) assumed distance. All values except the stellar mass, recessional velocity and half-light radius at our assumed distance are taken from \citet{Trujillo2021}.} 
    \label{tab:df9summary}
\end{table}


\section{Literature Data for DF9} \label{sec:imaging}
We note that DF9 was also presented in \citet{Trujillo2021} as SDSS J024007.01–081344.4. From their deep imaging, we take the Right Ascension, Declination, apparent $g-$band magnitude, central $g-$band surface brightness and $g-$band half-light radius. We list these in Table \ref{tab:df9summary}. We note that, based on their measured half-light radius and central surface brightness, DF9 is both too bright and too small to be considered an `ultra-diffuse galaxy' using the definition of \citet{vanDokkum2015}.

We additionally note the central compact object in DF9 that was identified as a possible nucleus from \citet{Buzzo2023}. They reported an apparent magnitude of $m_{g} = 21.9$~mag ($M_g = -9.6$~mag) and noted it has colours consistent with those of a GC ($\langle g-i \rangle = 0.71$~ and $\langle u-i \rangle = 1.68$). This is despite it being quite luminous for a `normal' GC.

\begin{figure}
    \centering
    \includegraphics[width = 0.48 \textwidth]{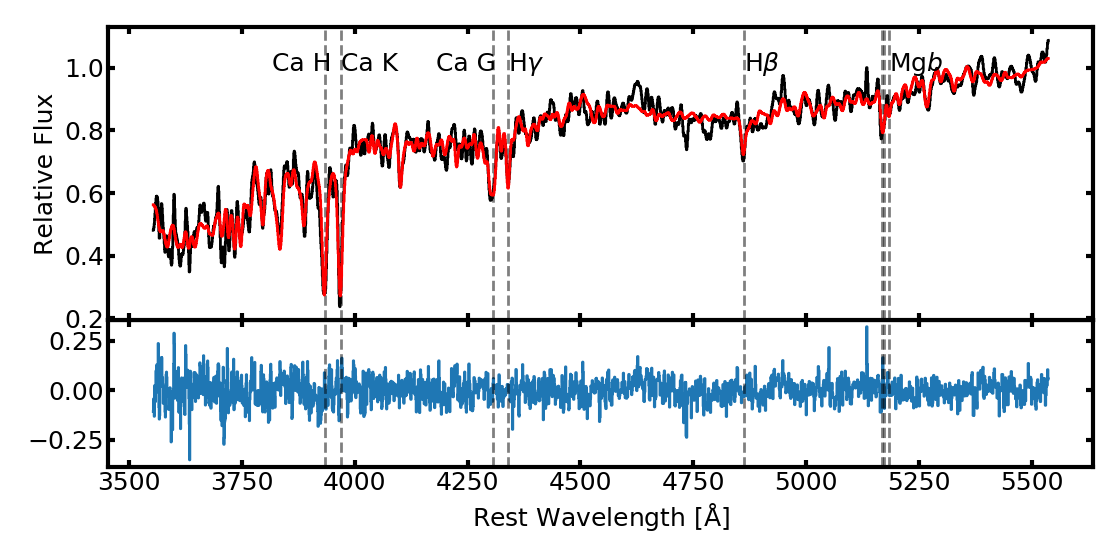}
    \caption{A Gaussian smoothed ($\sigma$ = 1 \AA) KCWI spectrum for DF9's stellar body (black; spectrum 2 in text) with example \texttt{pPXF} fit (red) displayed at the rest wavelengths. Residuals from the non-smoothed fit are shown at the bottom (blue). The prominent Hydrogen, Calcium and Magnesium absorption features are indicated by dashed vertical lines.}
    \label{fig:spectrum}
\end{figure}

\section{Stellar Population Results} \label{sec:results}

To extract stellar population parameters (i.e., mean mass-weighted ages and metallicities), we fit each spectrum with \texttt{pPXF} \citep{Cappellari2017} and the MILES stellar library \citep{Vazdekis2015}. We use MILES simple stellar population models assuming a Kroupa initial mass function \citep{Kroupa2001} and BaSTI isochrones. These models have total metallicities ranging from [$Z$/H] = $-$2.42 to +0.4 dex and ages ranging from 0.03 to 14 Gyr. The version of these models used assumes solar-scaled abundances (i.e., MILES BaseFe). 

The recessional velocity of DF9 was measured from a single fit with an additional 8th-order additive and multiplicative polynomial. From this fit, we measure a recessional velocity of $1680\pm10$ km s$^{-1}$ after barycentric correction \citep{Tollerud}. The uncertainty on the recessional velocity is that given by \texttt{pPXF} however we note it will not include uncertainty in our wavelength calibration which is likely $\sim$ 40 km s$^{-1}$. Our recessional velocity corrects the erroneous listing of DF9 as being a $z=0.933$ quasar from SDSS.

We then proceeded to measure the age and metallicity of DF9. These were measured as the median values of 256 fits resulting from all combinations of including a 0-15 degree additive and a 0-15 degree additive polynomial in the fitting procedure. Additive and multiplicative polynomials correct for continuum shape errors that may be introduced during flux calibration. Figure \ref{fig:spectrum} shows an example of one such \texttt{pPXF} fit to the spectrum of DF9's stellar body (spectrum 2). Uncertainties are taken from the 14th and 86th percentiles of the resulting parameter distributions. Our final results using this method for the mean mass-weighted age and metallicity of both spectra we extract are shown in Table \ref{tab:compare}. 


We checked for the effect of using the non-alpha-enhanced MILES models by refitting the data with the alpha-enhanced MILES model (i.e., [$\alpha$/Fe] = 0.4 dex]). The usage of alpha-enhanced models results in metallicities that are on average 0.2 dex more metal-rich and 0.57 Gyr older than the values quoted in Table \ref{tab:compare}. We choose, however, to quote values for the non-alpha enhanced models as these represent the best match to the methods used in \citet{Fensch2018} for DF2 which we use for comparison in the Discussion. We note all of our \texttt{pPXF} fitting does not apply regularisation to the output template weighting. Doing so will result in slightly younger ages ($\sim$ 1 Gyr depending on the choice of value). 

Based on our fitting of DF9, we measure a mass-to-light ratio in $g-$band of 2.2 which implies a total stellar mass for the galaxy of 1.31$\times10^8$~$\mathrm{M_\odot}$. We include this value in Table \ref{tab:df9summary}. This stellar mass is approximately that of DF4 and only slightly less than that of DF2. Based on our fitting of its nucleus alone, we measure a mass-to-light ratio in $g-$band of 1.9 which implies a total stellar for the nucleus of 1.46$\times10^6$~$\mathrm{M_\odot}$. Mass-to-light ratios assume the same \citet{Kroupa2001} initial mass function as our stellar library.


\begin{figure*}
    \centering
    \includegraphics[width = 0.95 \textwidth]{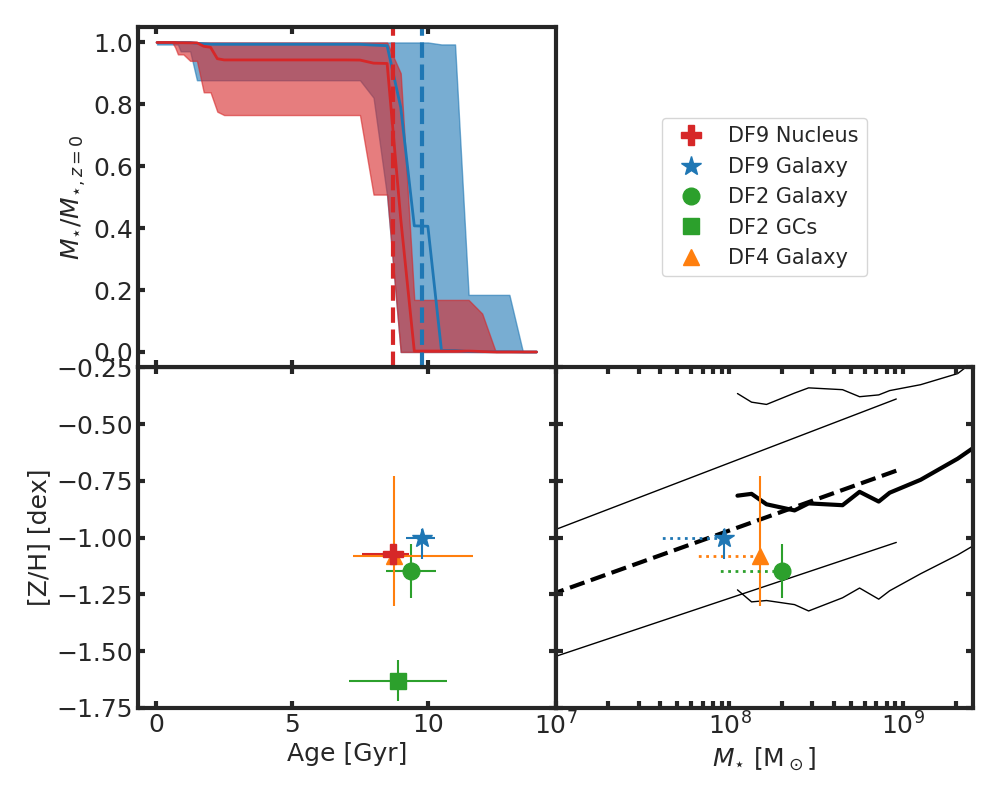}
    \caption{A comparison of our derived stellar population parameters of DF9 and its nucleus to measurements for DF2 and DF4. \textit{Top Panel:} \texttt{pPXF} derived, non-regularised star formation histories for both DF9 (blue) and its nucleus (red). Solid lines indicate the mean relation with shading between the earliest and latest star formation histories derived in our fitting. Vertical lines represent the mass-weighted ages of each. The star formation histories are consistent with both the nucleus and the galaxy forming at the same time. This conclusion will not change under reasonable regularisation choices. \textit{Bottom left:} Mass-weighted age \textit{vs.} metallicity. We show our values for DF9's nucleus (red plus) and for the stellar body without the nucleus (blue star). These are compared to: the mean age and metallicity of the stellar body of DF2 from \citet{Fensch2018}, \citet{RuizLara2019} and \citet[green circle]{Buzzo2022}; the GCs around DF2 from \citet[green square]{Fensch2018} and DF4's stellar body from \citet[orange triangle]{Buzzo2022}. The ages and metallicities of DF2, DF4 and DF9 are within uncertainties of one another but DF2's GC system is significantly more metal-poor. \textit{Bottom Right:} Metallicity \textit{vs.} stellar mass. Plotted points for the galaxies' stellar bodies are the same as in the previous panel. Mass--metallicity relationships are included based on \citet{Panter2008} and \citet{Simon2019}. The dotted lines indicate the expected change in stellar mass if the galaxies are at the closer claimed distance \citep{Trujillo2019}. All three galaxies are within the scatter of established mass metallicity relations. }
    \label{fig:corner}
\end{figure*}
\section{Discussion} \label{sec:discussion}

\subsection{Formation of the Nucleus}
In \citet{Fahrion2022} fig. 6 (and \citealp{Fahrion2020}; \citealp{Fahrion2021} therein), a transition is seen between galaxies with nuclear star clusters forming via GC mergers and in-situ nuclear star cluster formation. Based on the stellar mass of DF9 and its nucleus we expect it to reside in the regime where other galaxies are known to mostly have nuclei forming via GC mergers. However, the nucleus has approximately 7 times the average mass of a Milky Way GC ($\sim2\times10^{5}$~$\mathrm{M_{\odot}}$ \citealp{Harris1996}; 2010 version), suggesting a moderate number of GCs would be required if it were to form via GC mergers. This is at a slight tension with the relatively sparse GC system observed at present times ($\sim 3$~GCs; \citealp{Buzzo2023}), as it would require a large fraction of its historical GC system to have merged into the nucleus.

Furthermore, in the \textit{top panel} of Figure \ref{fig:corner} we show the star formation histories of DF9 and its nucleus. These star formation histories are not regularised, but our conclusions do not change under reasonable regularisation choices. Given their similar profiles, it seems quite likely that the nucleus formed at a similar epoch to the galaxy. This is further supported by the common metallicities between the two components suggesting that both formed from similarly enriched gas. This is not as expected in the GC merger formation scenario where the nucleus tends to be older and more metal-poor than the surrounding galaxy. It seems therefore likely that some combination of in-situ formation and GC-mergers took place in the formation of the nucleus of DF9.



\subsection{Formation of DF9}
In Figure \ref{fig:corner}, \textit{lower left}, we show the ages and metallicities we have derived for DF9 in comparison to results from the literature for DF2's stellar body, DF2's GC system and DF4's stellar body \citep{Fensch2018, RuizLara2019, Buzzo2022}. For DF2 we plot the error-weighted average of the three age/metallicity measurements from \citet{Fensch2018}, \citet{RuizLara2019} and \citet{Buzzo2022} so as not to crowd the plot. We find that the ages and metallicities of the three galaxies (i.e., DF2, DF4 and DF9) are within uncertainties of one another.

Despite the similar stellar populations, there exists the possibility that the three galaxies have unrelated formation pathways. In Figure \ref{fig:corner}, \textit{lower right}, we show DF2, DF4 and DF9 in stellar mass -- metallicity space. We include the established relationships of \citet{Panter2008} and \citet{Simon2019}. We note all of DF2/DF4/DF9 lie within the scatter of the mass -- metallicity relationship. The common metallicities of the three galaxies may, therefore, simply be a consequence of the known mass -- metallicity relationship. Therefore, we cannot be conclusive that the formation of these galaxies must be related, and that their similar stellar population properties are not the result of random chance. Despite this, for the remainder of the discussion, we will assume a common formation pathway for all three galaxies.

\begin{table}
    \centering
    \begin{tabular}{lllll}
    \hline
    Data & Object & {[}$Z$/H{]} {[}dex{]} & Age {[}Gyr{]} & Source \\ \hline
    Spectrum & DF9 & $-$1.00$^{+0.04}_{-0.09}$) & 9.79($^{+0.48}_{-0.6}$) & This work \\
    Spectrum & DF9 Nucleus & $-$1.07($^{+0.03}_{-0.04}$) & 8.72($^{+0.57}_{-1.16}$) & This work \\
    Spectrum & DF2 & $-$1.20 (0.07) & 9.8 (0.5) & RL19 \\
    Spectrum & DF2 & $-$1.07 (0.12) & 8.9 (1.5) & F19 \\
    Spectrum & DF2 GCs & $-$1.63 (0.09) & 8.9 (1.8) & F19 \\
    SED & DF2 & $-$1.11($^{+0.35}_{-0.28}$) & 7.97($^{+1.83}_{-2.73}$) & B22 \\
    Mean & DF2 & $-$1.14 (0.12) & 9.4 (0.9) & This work \\
    SED & DF4 & $-$1.08($^{+0.35}_{-0.22}$) & 8.76($^{+2.91}_{-1.51}$) & B22 \\ \hline
    \end{tabular}
    \caption{A comparison of the stellar population properties derived in this work for DF9 to those for DF2 and DF4 from the literature. From left to right the columns are: 1) the data used to derive the stellar populations, 2) the target galaxy, 3) [$Z$/H] metallicity, 4) Age in Gyr and 5) the source for the data. Where relevant, uncertainties are given in parentheses. The values quoted for DF9 are for a spectrum that does not include the nucleus. The row for DF2 listed as being the `Mean' presents the error-weighted average of three studies on DF2. RL19 refers to \citet{RuizLara2019}. F19 refers to \citet{Fensch2018}. B22 refers to \citet{Buzzo2022}.}
    \label{tab:compare}
\end{table}

\subsection{The Bullet Trail}
Having the ages and metallicities of DF2/DF4/DF9 within uncertainties of one another is as expected if they formed in the ``bullet dwarf'' formation scenario of vD+22. However, we show that in Figure \ref{fig:corner}, \textit{lower left}, the GC system of DF2 is on average more metal-poor than its stellar body \citep{Fensch2018}. Furthermore, given the similar colours of DF2's and DF4's GC systems and how the stellar bodies of both galaxies are on average redder than their GC systems \citep{vanDokkum2022b}, it seems likely DF4's GC system may also be more metal-poor than its stellar body. 

In the context of the bullet scenario, it is not clear how the galaxy and its GC system could have differing metallicity if they both form in a burst of high-pressure star formation from the same gas (which is needed to explain the overluminous GCs). Further star formation would be required to enrich the mass-weighted metallicity of the galaxy with respect to its GCs. However, this star formation would also be required to not have a large effect on the mass-weighted age that is measured to be within uncertainties of the age of the GCs. Further modelling of the post-collision metallicity enrichment of the galaxies is needed to understand this dilemma. 

A separate source for confirmation of this Bullet trail hypothesis would come from the recessional velocity of DF9. We calculate DF9 to be 14.3 arcminutes west of NGC~1052. Using the relation cz = 1700 $-$ 10 $\Delta {\rm RA}$ from vD+22 we predict a recessional velocity of 1557 km s$^{-1}$ for DF9. This is 123 km s$^{-1}$ (or $\sim12\times$ our uncertainty) lower than our measured recessional velocity (i.e., 1680 km s$^{-1}$). vD+22 noted that this relationship is expected to have considerable scatter and that the predicted recessional velocities based on galaxies' 2D projected positions may not be linearly increasing for most geometries once calculated in 3D. We suggest that, even in the case of considerable scatter, DF9 would not follow this relationship if it were linear as calculated. In order for the bullet formation scenario to be valid and cause the formation of DF9, the 3D geometry of the bullet trail must be such that the relationship is non-linear once projected into 2D. 

\subsection{Tidal Debris}
These galaxies may all have formed as tidal dwarf galaxies from gas stripped during a galaxy interaction, representing a more gentle case of the above ``bullet dwarf'' formation scenario. The common ages and metallicities are then also evidence for this tidal formation. Evidence from other systems suggests that galaxies of similar $R_{\rm e}$, stellar mass and low surface brightness to DF2/DF4/DF9 can form as tidal debris as they are embedded in tidal debris filaments (see e.g., \citealp{Roman2021} or \citealp{Iodice2021}).

We note that DF2 is not detected in deep HI imaging \citep{Sardone2019} and there are no convincing HI structures associated with any of DF2/DF4/DF9 \citep{Muller2019}. Were these galaxies to form from tidally stripped HI-gas, they must have now depleted their gas reservoir below the detection limits of these surveys. This is possible for these galaxies given their intermediate/old ages. However, in this generalised tidal debris scenario, it is difficult to explain the anomalous GC systems found for both DF2 and DF4 \citep{Shen2020}. For example, simulations of GC mergers based on the known observations of DF2 and DF4 demonstrate that these are not able to explain the over-luminous GC systems \citep{Dutta2020}. An additional special mechanism would therefore be required to induce the formation of the over-luminous GCs.

In addition to these problems, tidal dwarf galaxies are known to exhibit high metallicities for their stellar masses owing to their formation from enriched material (see e.g., \citealp{Sales2020} or \citealp{Roman2021}). All of our galaxies follow the established mass -- metallicity relationship and thus do not have the high metallicities expected by this scenario (Figure \ref{fig:corner}, \textit{lower right}). This conclusion is drawn largely irrespective of their distance. We show with a dotted line in Figure \ref{fig:corner}, \textit{lower right}, the $\sim2.5\times$ lower stellar mass that would be measured for the galaxies at a 13 Mpc distance \citep{Trujillo2019} and would still have them within the scatter of the established relationships. A possible explanation is that they could instead be tidal dwarfs formed at higher redshift where gas is less enriched (see e.g., \citealp{Fensch2019b} section 5.5). In this case, it would not be clear how these galaxies would survive the tidal field of the group until the present day due to their low dark matter content.


\subsection{Tidally Stripped Stellar Material}
It has also been proposed that DF2 and DF4 may each be the remnants of a tidally stripped galaxy \citep{Ogiya2018, Ogiya2021, Nusser2020, Montes2020, Maccio2021, Jackson2021, Moreno2022}. Notably, if this is the case there is no need for the galaxies to be stripped at the same time (and thus they wouldn't necessarily have similar stellar populations). Deep imaging from \citet{Muller2019} failed to find the low surface brightness features that may be expected, in this scenario, to be associated with each galaxy. The stellar ages for DF2/DF4/DF9 allow any such event to have occurred up to $\sim9$~Gyr ago. Therefore, any tidal features associated with the galaxies formation may have dissipated beyond the detection limit of \citet{Muller2019}. However, this does not exclude the galaxies from \textit{currently} undergoing a tidal interaction, for which there is evidence \citep{Montes2020, Keim2022}. 

In order to produce the currently observed large GC systems for DF2/DF4 ($\sim 19$~each; \citealp{Shen2020}), a tidal stripping scenario would require the progenitor to be unrealistically GC-rich \citep{Ogiya2021}. The subsequent stripping would be expected to leave a trail of stripped, intra-group GCs that are presently not observed in the group (\citealp{Buzzo2023}). It is difficult to envision a stripping scenario that removes large quantities of the stellar mass without removing some of the GC system. The only possibility for this formation hypothesis to be valid is that the progenitor had a GC system that was more compact than usual and hence more resistant to tidal stripping (see appendix B of \citealp{Ogiya2021}). 

\section{Conclusions} \label{sec:conclusions}
In this work, we have performed a stellar population analysis of Keck/KCWI data targeting the galaxy DF9 to investigate its formation history. We placed particular emphasis on comparing our stellar population results with those of the other nearby dwarf galaxies DF2 and DF4 around NGC~1052. Our main conclusions are as follows:

\begin{itemize}
    \item We measure a recessional velocity 1680 $\pm$ 10 km/s for DF9 which places the galaxy in the NGC~1052 group and corrects the erroneous listing of the galaxy as a high-redshift quasar.
    
    \item We recover an age and metallicity for the nucleus of DF9 that are in agreement with those of the stellar body of the galaxy. The high luminosity of the nucleus suggests a total mass that is approximately 7 times the average for a Milky Way GC. We, therefore, suggest that the nucleus of DF9 formed via a combination of GC mergers and/or in-situ star formation.
\end{itemize}
  
The age and metallicity we recover for the stellar body of DF9 are within uncertainties of literature ages and metallicities for the galaxies DF2 and DF4 also in the NGC~1052 group. Using them:
    
\begin{itemize}
    \item We investigate the likelihood that the commonality of characteristics of DF2/DF4/DF9 could be the result of random chance. The metallicities of the three galaxies may be the result of all following the dwarf stellar mass -- metallicity relation. Given their environment, it is difficult to be conclusive on the likelihood of their common age occurring randomly. 

    \item We investigate the recent proposal that these three galaxies may have formed as part of a ``bullet dwarf'' collision. Their common ages and metallicities are as expected for this scenario. However, the radial velocity of DF9 does not fit with a geometrically linear interpretation of the remnant dwarf trail. DF9 is either not part of the trail or the correlation of projected galaxy velocities cannot be linear. Furthermore differences between the metallicities of the GCs around DF2/DF4 and their stellar bodies require an explanation in the scenario as currently posed.
    
    \item We investigate a more general case of these galaxies forming from HI tidal debris. Our conclusions are the same as those for the ``bullet dwarf'' collision. However, it is now more difficult to explain the anomalous GC systems of two galaxies in the trail, DF2/DF4, without the special star-forming conditions induced by a high-speed collision. Additionally, the GCs' lack of an elevated metallicity for their stellar masses is at odds with this scenario.

    \item We investigate if these galaxies may have formed as the result of tidal stripping. Our intermediate age confirms that each interaction may have occurred sufficiently long ago to allow tidal features to dissipate below the surface brightness limits of current imaging. 
\end{itemize}

\section*{Acknowledgements}

We thank the referee for their careful, constructive review of our manuscript. We thank M. Keim and P. van Dokkum for conversations that aided the creation of this work. This research was supported by the Australian Research Council Centre of Excellence for All Sky Astrophysics in 3 Dimensions (ASTRO 3D), through project number CE170100013. AFM acknowledges funding from RYC2021-031099-I of MICIN/AEI/10.13039/501100011033/ UE. This project received funding from the Australian Research Council through DP220101863. This work was partially supported by a NASA Keck PI Data Award, administered by the NASA Exoplanet Science Institute. Some of the data presented herein were obtained at the W. M. Keck Observatory, which is operated as a scientific partnership among the California Institute of Technology, the University of California and the National Aeronautics and Space Administration. The Observatory was made possible by the generous financial support of the W. M. Keck Foundation. The authors wish to recognise and acknowledge the very significant cultural role and reverence that the summit of Maunakea has always had within the indigenous Hawaiian community.  We are most fortunate to have the opportunity to conduct observations from this mountain. 
\section*{Data Availability}
The KCWI data presented are available via the Keck Observatory Archive (KOA) 18 months after observations are taken.



\bibliographystyle{mnras}
\bibliography{bibliography} 

\begin{thebibliography}{}
\makeatletter
\relax
\def\mn@urlcharsother{\let\do\@makeother \do\$\do\&\do\#\do\^\do\_\do\%\do\~}
\def\mn@doi{\begingroup\mn@urlcharsother \@ifnextchar [ {\mn@doi@}
  {\mn@doi@[]}}
\def\mn@doi@[#1]#2{\def\@tempa{#1}\ifx\@tempa\@empty \href
  {http://dx.doi.org/#2} {doi:#2}\else \href {http://dx.doi.org/#2} {#1}\fi
  \endgroup}
\def\mn@eprint#1#2{\mn@eprint@#1:#2::\@nil}
\def\mn@eprint@arXiv#1{\href {http://arxiv.org/abs/#1} {{\tt arXiv:#1}}}
\def\mn@eprint@dblp#1{\href {http://dblp.uni-trier.de/rec/bibtex/#1.xml}
  {dblp:#1}}
\def\mn@eprint@#1:#2:#3:#4\@nil{\def\@tempa {#1}\def\@tempb {#2}\def\@tempc
  {#3}\ifx \@tempc \@empty \let \@tempc \@tempb \let \@tempb \@tempa \fi \ifx
  \@tempb \@empty \def\@tempb {arXiv}\fi \@ifundefined
  {mn@eprint@\@tempb}{\@tempb:\@tempc}{\expandafter \expandafter \csname
  mn@eprint@\@tempb\endcsname \expandafter{\@tempc}}}

\bibitem[\protect\citeauthoryear{{Buzzo} et~al.,}{{Buzzo}
  et~al.}{2022}]{Buzzo2022}
{Buzzo} M.~L.,  et~al., 2022, \mn@doi [\mnras] {10.1093/mnras/stac2442}, \href
  {https://ui.adsabs.harvard.edu/abs/2022MNRAS.517.2231B} {517, 2231}

\bibitem[\protect\citeauthoryear{{Buzzo}, {Forbes}, {Brodie}, {Janssens},
  {Couch}, {Romanowsky}  \& {Gannon}}{{Buzzo} et~al.}{2023}]{Buzzo2023}
{Buzzo} M.~L.,  {Forbes} D.~A.,  {Brodie} J.~P.,  {Janssens} S.~R.,  {Couch}
  W.~J.,  {Romanowsky} A.~J.,   {Gannon} J.~S.,  2023, \mn@doi [\mnras]
  {10.1093/mnras/stad1012}, \href
  {https://ui.adsabs.harvard.edu/abs/2023MNRAS.522..595B} {522, 595}

\bibitem[\protect\citeauthoryear{{Cappellari}}{{Cappellari}}{2017}]{Cappellari2017}
{Cappellari} M.,  2017, \mn@doi [\mnras] {10.1093/mnras/stw3020}, \href
  {https://ui.adsabs.harvard.edu/abs/2017MNRAS.466..798C} {466, 798}

\bibitem[\protect\citeauthoryear{{Danieli}, {van Dokkum}, {Conroy}, {Abraham}
  \& {Romanowsky}}{{Danieli} et~al.}{2019}]{Danieli2019}
{Danieli} S.,  {van Dokkum} P.,  {Conroy} C.,  {Abraham} R.,   {Romanowsky}
  A.~J.,  2019, \mn@doi [\apjl] {10.3847/2041-8213/ab0e8c}, \href
  {https://ui.adsabs.harvard.edu/abs/2019ApJ...874L..12D} {874, L12}

\bibitem[\protect\citeauthoryear{{Dutta Chowdhury}, {van den Bosch}  \& {van
  Dokkum}}{{Dutta Chowdhury} et~al.}{2020}]{Dutta2020}
{Dutta Chowdhury} D.,  {van den Bosch} F.~C.,   {van Dokkum} P.,  2020, \mn@doi
  [\apj] {10.3847/1538-4357/abb947}, \href
  {https://ui.adsabs.harvard.edu/abs/2020ApJ...903..149D} {903, 149}

\bibitem[\protect\citeauthoryear{{Emsellem} et~al.,}{{Emsellem}
  et~al.}{2019}]{Emsellem2019}
{Emsellem} E.,  et~al., 2019, \mn@doi [\aap] {10.1051/0004-6361/201834909},
  \href {https://ui.adsabs.harvard.edu/abs/2019A&A...625A..76E} {625, A76}

\bibitem[\protect\citeauthoryear{{Fahrion} et~al.,}{{Fahrion}
  et~al.}{2020}]{Fahrion2020}
{Fahrion} K.,  et~al., 2020, \mn@doi [\aap] {10.1051/0004-6361/201937120},
  \href {https://ui.adsabs.harvard.edu/abs/2020A&A...634A..53F} {634, A53}

\bibitem[\protect\citeauthoryear{{Fahrion} et~al.,}{{Fahrion}
  et~al.}{2021}]{Fahrion2021}
{Fahrion} K.,  et~al., 2021, \mn@doi [\aap] {10.1051/0004-6361/202140644},
  \href {https://ui.adsabs.harvard.edu/abs/2021A&A...650A.137F} {650, A137}

\bibitem[\protect\citeauthoryear{{Fahrion} et~al.,}{{Fahrion}
  et~al.}{2022}]{Fahrion2022}
{Fahrion} K.,  et~al., 2022, \mn@doi [\aap] {10.1051/0004-6361/202244932},
  \href {https://ui.adsabs.harvard.edu/abs/2022A&A...667A.101F} {667, A101}

\bibitem[\protect\citeauthoryear{{Fensch} et~al.,}{{Fensch}
  et~al.}{2019a}]{Fensch2018}
{Fensch} J.,  et~al., 2019a, \mn@doi [\aap] {10.1051/0004-6361/201834911},
  \href {https://ui.adsabs.harvard.edu/abs/2019A&A...625A..77F} {625, A77}

\bibitem[\protect\citeauthoryear{{Fensch} et~al.,}{{Fensch}
  et~al.}{2019b}]{Fensch2019b}
{Fensch} J.,  et~al., 2019b, \mn@doi [\aap] {10.1051/0004-6361/201834403},
  \href {https://ui.adsabs.harvard.edu/abs/2019A&A...628A..60F} {628, A60}

\bibitem[\protect\citeauthoryear{{Fosbury}, {Mebold}, {Goss}  \&
  {Dopita}}{{Fosbury} et~al.}{1978}]{Fosbury1978}
{Fosbury} R.~A.~E.,  {Mebold} U.,  {Goss} W.~M.,   {Dopita} M.~A.,  1978,
  \mn@doi [\mnras] {10.1093/mnras/183.4.549}, \href
  {https://ui.adsabs.harvard.edu/abs/1978MNRAS.183..549F} {183, 549}

\bibitem[\protect\citeauthoryear{{Gannon}, {Forbes}, {Romanowsky},
  {Ferr{\'e}-Mateu}, {Couch}  \& {Brodie}}{{Gannon} et~al.}{2020}]{Gannon2020}
{Gannon} J.~S.,  {Forbes} D.~A.,  {Romanowsky} A.~J.,  {Ferr{\'e}-Mateu} A.,
  {Couch} W.~J.,   {Brodie} J.~P.,  2020, \mn@doi [\mnras]
  {10.1093/mnras/staa1282}, \href
  {https://ui.adsabs.harvard.edu/abs/2020MNRAS.495.2582G} {495, 2582}

\bibitem[\protect\citeauthoryear{{Harris}}{{Harris}}{1996}]{Harris1996}
{Harris} W.~E.,  1996, \mn@doi [\aj] {10.1086/118116}, \href
  {https://ui.adsabs.harvard.edu/abs/1996AJ....112.1487H} {112, 1487}

\bibitem[\protect\citeauthoryear{{Iodice} et~al.,}{{Iodice}
  et~al.}{2021}]{Iodice2021}
{Iodice} E.,  et~al., 2021, \mn@doi [\aap] {10.1051/0004-6361/202141086}, \href
  {https://ui.adsabs.harvard.edu/abs/2021A&A...652L..11I} {652, L11}

\bibitem[\protect\citeauthoryear{{Jackson} et~al.,}{{Jackson}
  et~al.}{2021}]{Jackson2021}
{Jackson} R.~A.,  et~al., 2021, \mn@doi [\mnras] {10.1093/mnras/stab093}, \href
  {https://ui.adsabs.harvard.edu/abs/2021MNRAS.502.1785J} {502, 1785}

\bibitem[\protect\citeauthoryear{{Karachentsev}, {Karachentseva}, {Suchkov}  \&
  {Grebel}}{{Karachentsev} et~al.}{2000}]{Karachentsev2000}
{Karachentsev} I.~D.,  {Karachentseva} V.~E.,  {Suchkov} A.~A.,   {Grebel}
  E.~K.,  2000, \mn@doi [\aaps] {10.1051/aas:2000249}, \href
  {https://ui.adsabs.harvard.edu/abs/2000A&AS..145..415K} {145, 415}

\bibitem[\protect\citeauthoryear{{Keim} et~al.,}{{Keim}
  et~al.}{2022}]{Keim2022}
{Keim} M.~A.,  et~al., 2022, \mn@doi [\apj] {10.3847/1538-4357/ac7dab}, \href
  {https://ui.adsabs.harvard.edu/abs/2022ApJ...935..160K} {935, 160}

\bibitem[\protect\citeauthoryear{{Kroupa}}{{Kroupa}}{2001}]{Kroupa2001}
{Kroupa} P.,  2001, \mn@doi [\mnras] {10.1046/j.1365-8711.2001.04022.x}, \href
  {https://ui.adsabs.harvard.edu/abs/2001MNRAS.322..231K} {322, 231}

\bibitem[\protect\citeauthoryear{{Lee}, {Shin}  \& {Kim}}{{Lee}
  et~al.}{2021}]{Lee2021}
{Lee} J.,  {Shin} E.-j.,   {Kim} J.-h.,  2021, \mn@doi [\apjl]
  {10.3847/2041-8213/ac16e0}, \href
  {https://ui.adsabs.harvard.edu/abs/2021ApJ...917L..15L} {917, L15}

\bibitem[\protect\citeauthoryear{{Macci{\`o}}, {Prats}, {Dixon}, {Buck},
  {Waterval}, {Arora}, {Courteau}  \& {Kang}}{{Macci{\`o}}
  et~al.}{2021}]{Maccio2021}
{Macci{\`o}} A.~V.,  {Prats} D.~H.,  {Dixon} K.~L.,  {Buck} T.,  {Waterval} S.,
   {Arora} N.,  {Courteau} S.,   {Kang} X.,  2021, \mn@doi [\mnras]
  {10.1093/mnras/staa3716}, \href
  {https://ui.adsabs.harvard.edu/abs/2021MNRAS.501..693M} {501, 693}

\bibitem[\protect\citeauthoryear{{Montes}, {Infante-Sainz}, {Madrigal-Aguado},
  {Rom{\'a}n}, {Monelli}, {Borlaff}  \& {Trujillo}}{{Montes}
  et~al.}{2020}]{Montes2020}
{Montes} M.,  {Infante-Sainz} R.,  {Madrigal-Aguado} A.,  {Rom{\'a}n} J.,
  {Monelli} M.,  {Borlaff} A.~S.,   {Trujillo} I.,  2020, \mn@doi [\apj]
  {10.3847/1538-4357/abc340}, \href
  {https://ui.adsabs.harvard.edu/abs/2020ApJ...904..114M} {904, 114}

\bibitem[\protect\citeauthoryear{{Moreno} et~al.,}{{Moreno}
  et~al.}{2022}]{Moreno2022}
{Moreno} J.,  et~al., 2022, \mn@doi [Nature Astronomy]
  {10.1038/s41550-021-01598-4}, \href
  {https://ui.adsabs.harvard.edu/abs/2022NatAs...6..496M} {6, 496}

\bibitem[\protect\citeauthoryear{{Morrissey} et~al.,}{{Morrissey}
  et~al.}{2018}]{Morrissey2018}
{Morrissey} P.,  et~al., 2018, \mn@doi [\apj] {10.3847/1538-4357/aad597}, \href
  {https://ui.adsabs.harvard.edu/abs/2018ApJ...864...93M} {864, 93}

\bibitem[\protect\citeauthoryear{{M{\"u}ller} et~al.,}{{M{\"u}ller}
  et~al.}{2019}]{Muller2019}
{M{\"u}ller} O.,  et~al., 2019, \mn@doi [\aap] {10.1051/0004-6361/201935463},
  \href {https://ui.adsabs.harvard.edu/abs/2019A&A...624L...6M} {624, L6}

\bibitem[\protect\citeauthoryear{{Nusser}}{{Nusser}}{2020}]{Nusser2020}
{Nusser} A.,  2020, \mn@doi [\apj] {10.3847/1538-4357/ab792c}, \href
  {https://ui.adsabs.harvard.edu/abs/2020ApJ...893...66N} {893, 66}

\bibitem[\protect\citeauthoryear{{Ogiya}}{{Ogiya}}{2018}]{Ogiya2018}
{Ogiya} G.,  2018, \mn@doi [\mnras] {10.1093/mnrasl/sly138}, \href
  {https://ui.adsabs.harvard.edu/abs/2018MNRAS.480L.106O} {480, L106}

\bibitem[\protect\citeauthoryear{{Ogiya}, {van den Bosch}  \&
  {Burkert}}{{Ogiya} et~al.}{2022}]{Ogiya2021}
{Ogiya} G.,  {van den Bosch} F.~C.,   {Burkert} A.,  2022, \mn@doi [\mnras]
  {10.1093/mnras/stab3658}, \href
  {https://ui.adsabs.harvard.edu/abs/2022MNRAS.510.2724O} {510, 2724}

\bibitem[\protect\citeauthoryear{{Panter}, {Jimenez}, {Heavens}  \&
  {Charlot}}{{Panter} et~al.}{2008}]{Panter2008}
{Panter} B.,  {Jimenez} R.,  {Heavens} A.~F.,   {Charlot} S.,  2008, \mn@doi
  [\mnras] {10.1111/j.1365-2966.2008.13981.x}, \href
  {https://ui.adsabs.harvard.edu/abs/2008MNRAS.391.1117P} {391, 1117}

\bibitem[\protect\citeauthoryear{{Rom{\'a}n}, {Jones}, {Montes},
  {Verdes-Montenegro}, {Garrido}  \& {S{\'a}nchez}}{{Rom{\'a}n}
  et~al.}{2021a}]{Roman2021}
{Rom{\'a}n} J.,  {Jones} M.~G.,  {Montes} M.,  {Verdes-Montenegro} L.,
  {Garrido} J.,   {S{\'a}nchez} S.,  2021a, \mn@doi [\aap]
  {10.1051/0004-6361/202141001}, \href
  {https://ui.adsabs.harvard.edu/abs/2021A&A...649L..14R} {649, L14}

\bibitem[\protect\citeauthoryear{{Rom{\'a}n}, {Castilla}  \&
  {Pascual-Granado}}{{Rom{\'a}n} et~al.}{2021b}]{Roman2021b}
{Rom{\'a}n} J.,  {Castilla} A.,   {Pascual-Granado} J.,  2021b, \mn@doi [\aap]
  {10.1051/0004-6361/202142161}, \href
  {https://ui.adsabs.harvard.edu/abs/2021A&A...656A..44R} {656, A44}

\bibitem[\protect\citeauthoryear{{Ruiz-Lara} et~al.,}{{Ruiz-Lara}
  et~al.}{2019}]{RuizLara2019}
{Ruiz-Lara} T.,  et~al., 2019, \mn@doi [\mnras] {10.1093/mnras/stz1237}, \href
  {https://ui.adsabs.harvard.edu/abs/2019MNRAS.486.5670R} {486, 5670}

\bibitem[\protect\citeauthoryear{{Sales}, {Navarro}, {Pe{\~n}afiel}, {Peng},
  {Lim}  \& {Hernquist}}{{Sales} et~al.}{2020}]{Sales2020}
{Sales} L.~V.,  {Navarro} J.~F.,  {Pe{\~n}afiel} L.,  {Peng} E.~W.,  {Lim} S.,
   {Hernquist} L.,  2020, \mn@doi [\mnras] {10.1093/mnras/staa854}, \href
  {https://ui.adsabs.harvard.edu/abs/2020MNRAS.494.1848S} {494, 1848}

\bibitem[\protect\citeauthoryear{{Sardone}, {Pisano}, {Burke-Spolaor},
  {Mascoop}  \& {Pol}}{{Sardone} et~al.}{2019}]{Sardone2019}
{Sardone} A.,  {Pisano} D.~J.,  {Burke-Spolaor} S.,  {Mascoop} J.~L.,   {Pol}
  N.,  2019, \mn@doi [\apjl] {10.3847/2041-8213/ab0084}, \href
  {https://ui.adsabs.harvard.edu/abs/2019ApJ...871L..31S} {871, L31}

\bibitem[\protect\citeauthoryear{{Shen}, {van Dokkum}  \& {Danieli}}{{Shen}
  et~al.}{2021a}]{Shen2020}
{Shen} Z.,  {van Dokkum} P.,   {Danieli} S.,  2021a, \mn@doi [\apj]
  {10.3847/1538-4357/abdd29}, \href
  {https://ui.adsabs.harvard.edu/abs/2021ApJ...909..179S} {909, 179}

\bibitem[\protect\citeauthoryear{{Shen} et~al.,}{{Shen}
  et~al.}{2021b}]{Shen2021}
{Shen} Z.,  et~al., 2021b, \mn@doi [\apjl] {10.3847/2041-8213/ac0335}, \href
  {https://ui.adsabs.harvard.edu/abs/2021ApJ...914L..12S} {914, L12}

\bibitem[\protect\citeauthoryear{{Shin}, {Jung}, {Kwon}, {Kim}, {Lee}, {Jo}  \&
  {Oh}}{{Shin} et~al.}{2020}]{Shin2020}
{Shin} E.-j.,  {Jung} M.,  {Kwon} G.,  {Kim} J.-h.,  {Lee} J.,  {Jo} Y.,   {Oh}
  B.~K.,  2020, \mn@doi [\apj] {10.3847/1538-4357/aba434}, \href
  {https://ui.adsabs.harvard.edu/abs/2020ApJ...899...25S} {899, 25}

\bibitem[\protect\citeauthoryear{{Silk}}{{Silk}}{2019}]{Silk2019}
{Silk} J.,  2019, \mn@doi [\mnras] {10.1093/mnrasl/slz090}, \href
  {https://ui.adsabs.harvard.edu/abs/2019MNRAS.488L..24S} {488, L24}

\bibitem[\protect\citeauthoryear{{Simon}}{{Simon}}{2019}]{Simon2019}
{Simon} J.~D.,  2019, \mn@doi [\araa] {10.1146/annurev-astro-091918-104453},
  \href {https://ui.adsabs.harvard.edu/abs/2019ARA&A..57..375S} {57, 375}

\bibitem[\protect\citeauthoryear{Tollerud}{Tollerud}{2015}]{Tollerud}
Tollerud E.,  2015, \url {https://gist.github.com/eteq/5000843}

\bibitem[\protect\citeauthoryear{{Trujillo-Gomez}, {Kruijssen}, {Keller}  \&
  {Reina-Campos}}{{Trujillo-Gomez} et~al.}{2021}]{TrujilloGomez2021}
{Trujillo-Gomez} S.,  {Kruijssen} J.~M.~D.,  {Keller} B.~W.,   {Reina-Campos}
  M.,  2021, \mn@doi [\mnras] {10.1093/mnras/stab1895}, \href
  {https://ui.adsabs.harvard.edu/abs/2021MNRAS.506.4841T} {506, 4841}

\bibitem[\protect\citeauthoryear{{Trujillo} et~al.,}{{Trujillo}
  et~al.}{2019}]{Trujillo2019}
{Trujillo} I.,  et~al., 2019, \mn@doi [\mnras] {10.1093/mnras/stz771}, \href
  {https://ui.adsabs.harvard.edu/abs/2019MNRAS.486.1192T} {486, 1192}

\bibitem[\protect\citeauthoryear{{Trujillo} et~al.,}{{Trujillo}
  et~al.}{2021}]{Trujillo2021}
{Trujillo} I.,  et~al., 2021, \mn@doi [\aap]
  {10.1051/0004-6361/20214160310.48550/arXiv.2109.07478}, \href
  {https://ui.adsabs.harvard.edu/abs/2021A&A...654A..40T} {654, A40}

\bibitem[\protect\citeauthoryear{{Vazdekis} et~al.,}{{Vazdekis}
  et~al.}{2015}]{Vazdekis2015}
{Vazdekis} A.,  et~al., 2015, \mn@doi [\mnras] {10.1093/mnras/stv151}, \href
  {https://ui.adsabs.harvard.edu/abs/2015MNRAS.449.1177V} {449, 1177}

\bibitem[\protect\citeauthoryear{{van Dokkum}, {Abraham}, {Merritt}, {Zhang},
  {Geha}  \& {Conroy}}{{van Dokkum} et~al.}{2015}]{vanDokkum2015}
{van Dokkum} P.~G.,  {Abraham} R.,  {Merritt} A.,  {Zhang} J.,  {Geha} M.,
  {Conroy} C.,  2015, \mn@doi [\apjl] {10.1088/2041-8205/798/2/L45}, \href
  {https://ui.adsabs.harvard.edu/abs/2015ApJ...798L..45V} {798, L45}

\bibitem[\protect\citeauthoryear{{van Dokkum} et~al.,}{{van Dokkum}
  et~al.}{2018a}]{vanDokkum2018}
{van Dokkum} P.,  et~al., 2018a, \mn@doi [\nat] {10.1038/nature25767}, \href
  {https://ui.adsabs.harvard.edu/abs/2018Natur.555..629V} {555, 629}

\bibitem[\protect\citeauthoryear{{van Dokkum} et~al.,}{{van Dokkum}
  et~al.}{2018b}]{vanDokkum2018c}
{van Dokkum} P.,  et~al., 2018b, \mn@doi [\apjl] {10.3847/2041-8213/aab60b},
  \href {https://ui.adsabs.harvard.edu/abs/2018ApJ...856L..30V} {856, L30}

\bibitem[\protect\citeauthoryear{{van Dokkum}, {Danieli}, {Abraham}, {Conroy}
  \& {Romanowsky}}{{van Dokkum} et~al.}{2019}]{vanDokkum2019}
{van Dokkum} P.,  {Danieli} S.,  {Abraham} R.,  {Conroy} C.,   {Romanowsky}
  A.~J.,  2019, \mn@doi [\apjl] {10.3847/2041-8213/ab0d92}, \href
  {https://ui.adsabs.harvard.edu/abs/2019ApJ...874L...5V} {874, L5}

\bibitem[\protect\citeauthoryear{{van Dokkum} et~al.,}{{van Dokkum}
  et~al.}{2022a}]{vanDokkum2022}
{van Dokkum} P.,  et~al., 2022a, \mn@doi [\nat] {10.1038/s41586-022-04665-6},
  \href {https://ui.adsabs.harvard.edu/abs/2022Natur.605..435V} {605, 435}

\bibitem[\protect\citeauthoryear{{van Dokkum} et~al.,}{{van Dokkum}
  et~al.}{2022b}]{vanDokkum2022b}
{van Dokkum} P.,  et~al., 2022b, \mn@doi [\apjl] {10.3847/2041-8213/ac94d6},
  \href {https://ui.adsabs.harvard.edu/abs/2022ApJ...940L...9V} {940, L9}

\makeatother
\end{thebibliography}








\bsp	
\label{lastpage}
\end{document}